# DEMO enhanced BPMN

Sérgio Guerreiro[1,2], Jan Dietz[3]

[1] INESC-ID, Rua Alves Redol 9, 1000-029 Lisbon, Portugal
[2] Instituto Superior Técnico, University of Lisbon, Lisbon, Portugal
[3] Delft University of Technology, Netherlands

Corresponding author: sergio.guerreiro@tecnico.ulisboa.pt
sapio-C4E@outlook.com

**Abstract**

*This paper presents an integration between DEMO (Design and Engineering Methodology for Organizations) and BPMN (Business Process Model and Notation). While BPMN is widely used for its intuitive, flow-based representation of business processes, it suffers from a lack of formal semantics, ambiguity, and limitations in modeling multi-party collaborations. In contrast, DEMO offers a theoretically robust, ontology-driven framework that focuses on abstracting the essential structure of business processes. A novel approach combining the rigor of DEMO's transaction patterns with the more practical, widely adopted BPMN framework is proposed and demonstrated. This integration allows for the benefits of DEMO's theoretical foundations to be utilized within BPMN diagrams, providing a more comprehensive and precise understanding of business processes. We argue that this combination enriches the modeling of business processes, providing a more coherent and reliable tool for both practitioners and researchers.*

## 1 Introduction

In the book titled "Enterprise Ontology - a human-centric approach to understanding the essence of organisation" [1], a bunch of theories is presented that together constitute the mental 'glasses' through which one 'sees' the ontological essence of an enterprise. The book also discusses the methodology DEMO (Design and Engineering Methodology for Organisations) for producing the so-called *essential model* of an enterprise. Abstracting in several ways from the every-day practice in an enterprise, this model provides a comprehensive, consistent, coherent, and concise understanding of an enterprise's ontological essence. It includes the understanding of the distinct *business processes* that take place in the operating enterprise. This understanding is primarily based on the inherent tree structures in business processes, next to understanding processes as flows, i.e. sequences of activities and events.

BPMN (Business Process Model and Notation) is a well-known approach to modelling business processes. It comprises an intuitive and flow-based way of understanding business processes, and of expressing them in diagrams. BPMN models can also be extended for the purpose of being executed [2]. The lack of clearly defined semantics leads to a lot of confusion in the practical application of BPMN [3]. In addition, there is a limited potential for verification, and the inability to model multi-party collaborations [4]. Although the latest release (BPMN 2.0.2)[1] addresses the issue of formal semantics to some extent, a number of challenges remain. Moreover, this version of BPMN comprises about 100 different symbols, which makes its practical application quite less 'intuitive' than originally envisioned.

Despite these drawbacks, BPMN is the de facto standard in business process modelling for professionals and researchers. In particular 'business people' appreciate it. They like to make models without much ado concerning precision and ambiguity. In terms of the distinction between the function and the construction perspective on enterprises [1], they pursue only the function perspective (the business of the enterprise) in contrast to the construction perspective (its organisation). Yet, the focus of BPMN is (implicitly though) on the construction perspective. Moreover, BPMN does not abstract from implementation in order to focus on ontology, nor does it distinguish between original, informational, and documental production [1].

In this short paper, we present a way to combine DEMO and BPMN so that one can benefit from the theoretical strengths of DEMO when using BPMN. After a short introduction to DEMO, in Sect. 2, and to BPMN, in Sect. 3, we discuss how the organisational building block in DEMO can be expressed in BPMN diagrams in Sect. 4. In Sect. 5, we discuss the practical use of the BPMN building blocks. Sect. 6 contains our reflections and conclusions. A basic knowledge of DEMO and BPMN is assumed.

## 2 DEMO

The building block of business processes in DEMO is the *transactor role*. It is the combination of a *transaction kind* and its executing *actor role* [1]. The process of every transaction, regardless its transaction kind, is a path through the Complete Transaction Pattern (CTP), which is shown in Fig. 1. Two actors are involved: one as the *initiator* of the transaction and one as the *executor*. The diagram looks very much like a Petrinet [5]. It has a richter semantics, however. Coordination events (C-events) are represented by disks, coordination acts (C-acts) by boxes. The grey coloured box with a diamond inside it, represents the production act and its resulting product. An arrow from a disk to a box represents a response link: the act is performed in response to the event. A straight line from a box to a disk represents a causal link: the occurrence of the event is the effect of performing the act.

The CTP consists of the *standard pattern* (in the middle) and the four *revocation patterns* (in the corners). The light grey-lined boxes indicate the responsibility areas of the initiator and the executor. The green path in Fig. 1 represents the success path or happy flow, in DEMO also called the *basic pattern*. It is next sequence of C-events and C-acts: (in), [rq], (rq), [pm], (pm), [da], (da), [ac], (ac). Note that we have skipped the production act, since the result of this act is not directly knowable.

---

[1] https://www.omg.org/spec/BPMN/2.0.2/PDF

By connecting transactor roles through initiator links, tree structures of transactor roles emerge. Every tree represents a *business process kind* in reality.

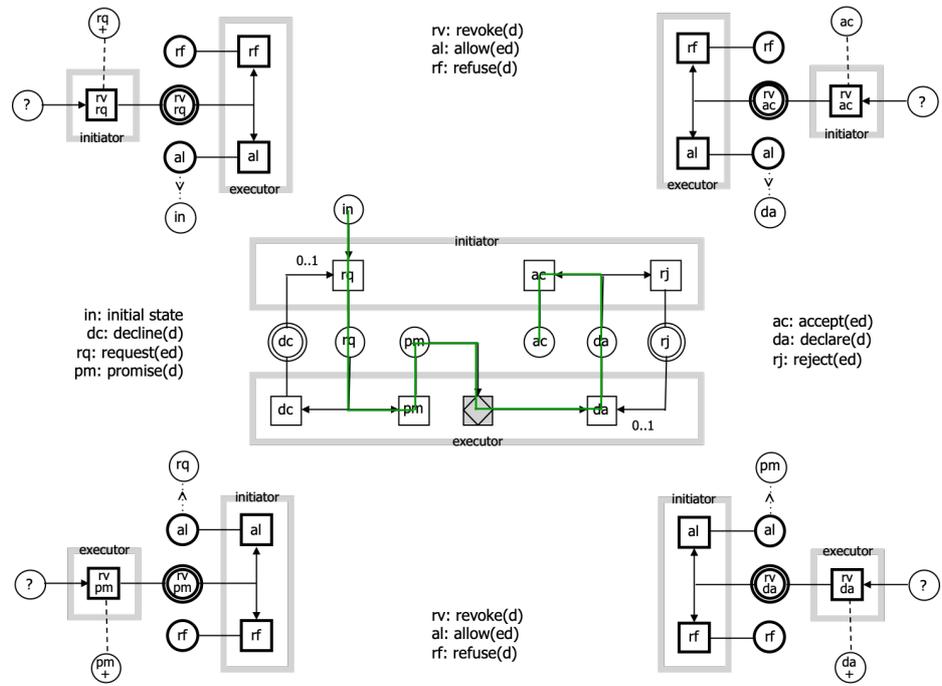

**Fig.1. The Complete Transaction Pattern in DEMO**

## 3 BPMN

In BPMN, a business process is conceived as a sequence of activities and events. The most basic modelling elements and their notation are listed in the Fig. 2. In the next sections, we will only make use of the symbols in this figure.

| Element | Description | Notation |
|---|---|---|
| Event | An Event is something that "happens" during the course of a Process (see page 238) or a Choreography (see page 339). These Events affect the flow of the model and usually have a cause (*trigger*) or an impact (*result*). Events are circles with open centers to allow internal markers to differentiate different *triggers* or *results*. There are three types of Events, based on when they affect the flow: Start, Intermediate, and End. | 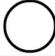 |
| Activity | An Activity is a generic term for work that company performs (see page 151) in a Process. An Activity can be atomic or non-atomic (compound). The types of Activities that are a part of a Process Model are: Sub-Process and Task, which are rounded rectangles. Activities are used in both standard Processes and in Choreographies. | 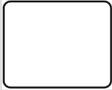 |
| Gateway | A Gateway is used to control the divergence and convergence of Sequence Flows in a Process (see page 145) and in a Choreography (see page 344). Thus, it will determine branching, forking, merging, and joining of paths. Internal markers will indicate the type of behavior control. | 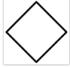 |
| Sequence Flow | A Sequence Flow is used to show the order that Activities will be performed in a Process (see page 97) and in a Choreography (see page 320). | 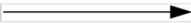 |
| Message Flow | A Message Flow is used to show the flow of Messages between two *Participants* that are prepared to send and receive them (see page 120). In BPMN, two separate Pools in a Collaboration Diagram will represent the two *Participants* (e.g., PartnerEntities and/or PartnerRoles). | 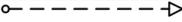 |
| Pool | A Pool is the graphical representation of a *Participant* in a Collaboration (see page 112). It also acts as a "swimlane" and a graphical container for partitioning a set of Activities from other Pools, usually in the context of B2B situations. A Pool MAY have internal details, in the form of the Process that will be executed. Or a Pool MAY have no internal details, i.e., it can be a "black box." | 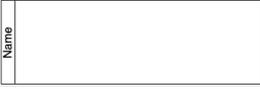 |
| Lane | A Lane is a sub-partition within a Process, sometimes within a Pool, and will extend the entire length of the Process, either vertically or horizontally (see on page 305). Lanes are used to organize and categorize Activities. | 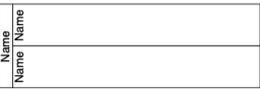 |
| Data Object | Data Objects provide information about what Activities require to be performed and/or what they produce (see page 205), Data Objects can represent a singular object or a collection of objects. Data Input and Data Output provide the same information for Processes. | 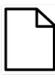 |
| Message | A Message is used to depict the contents of a communication between two *Participants* (as defined by a business PartnerRole or a business PartnerEntity—see on page 93). | 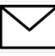 |

**Fig. 2.** The most basic elements of a BPMN diagram (excerpt from BPMN2.0 specification)

## 4 Expressing DEMO transactions in BPMN diagrams

In a BPMN diagram, one expresses basically the activities and events that can be observed. Consequently, tacitly performed coordination acts and their resulting events [1] will not appear in the diagram. In addition, 'exceptions', like the decline and reject in the standard pattern of DEMO (cf. Fig. 1) and all steps in the revocation patterns, will normally not appear in BPMN diagrams. Therefore, we need to express the basic, the standard, and the complete DEMO transaction pattern in BPMN diagrams. The resulting diagrams can then be used as 'building blocks' in BPMN models.

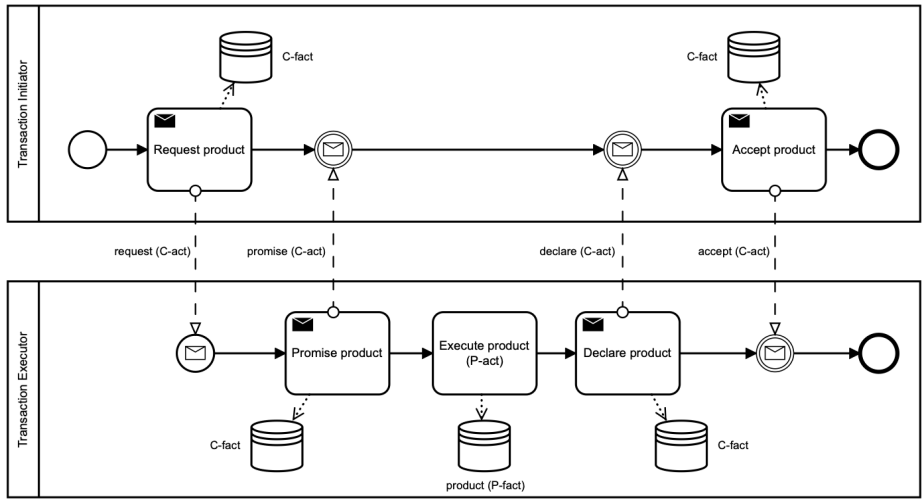

**Fig. 3. BPMN diagram of the basic DEMO transaction pattern**

Fig. 3 and Fig. 4 exhibit the expression in BPMN of respectively the basic and the standard transaction pattern. They are slightly modified versions of Figs. 5 and 6 in [6]. As for the complete transaction pattern, the corresponding BPMN diagram is too large to include in this short paper. However, one can find it in [6] and [7]. A pseudo BPMN diagram of the CTP is shown in Fig. 5. It is a copy of Fig. 3 in [6]. The next explanation holds. From any state (or: in response to any event), the initiator of a transaction may revoke her/his accept act (cf. Fig. 1). This is indicated by the purple circle and the purple arrow in the middle of Fig. 5. If the revoke is allowable, then the accept act is reversed (indicated by the activity Accept$^{-1}$), after which the process ends up in the state rejected (cf. Fig. 1). From there, the executor has the option to perform an adapted declare, but he/she can also decide to revoke her/his declare.

Analogous reasonings hold for the other three revocation kinds, indicated by the green (declare), the red (promise) and the blue (request) coloured circles and flows. All symbols in black constitute the basic transaction pattern (cf. Fig. 3) but in a simplified way.

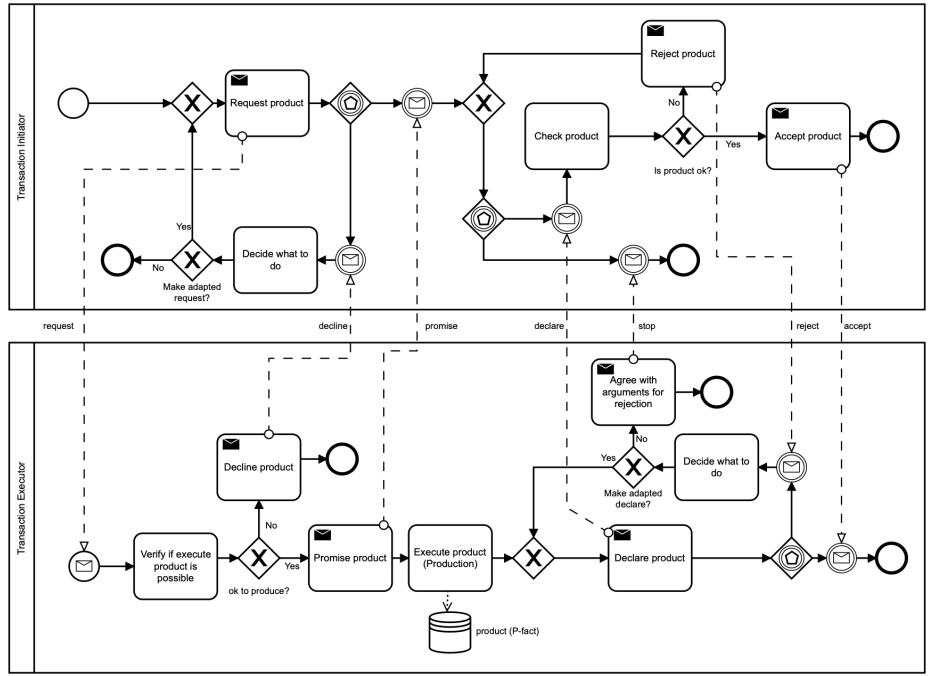

**Fig. 4.** BPMN diagram of the standard DEMO transaction pattern

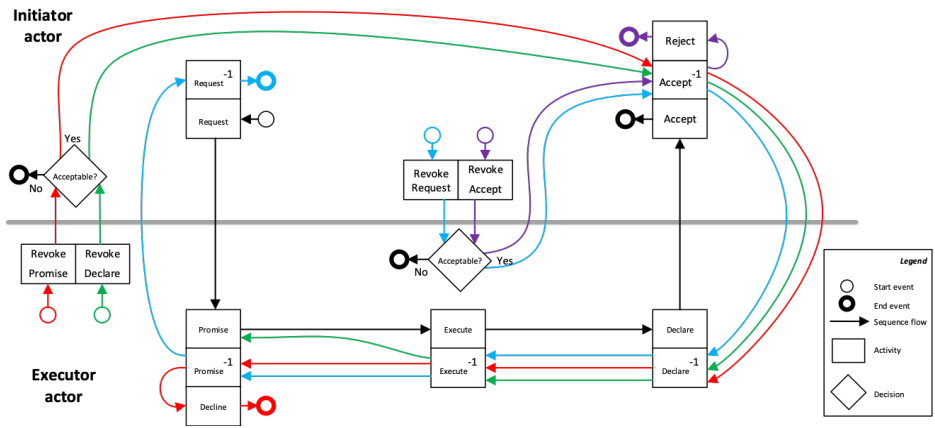

**Fig. 5.** Pseudo BPMN diagram of the complete DEMO transaction pattern.

## 5 Application of the building blocks

In this section we present an example of how the BPMN building blocks that we discussed in Sect. 4, can be applied to real cases, in order to illustrate their practical use.

The example is the case PoliGyn, as discussed in [1], Chap. 17. In Fig. 6, a part of the CSD (Coordination Structure Diagram) and the corresponding part of the PSD (Process Structure Diagram) are shown. From the state promised in a transaction TK01 (patient problem diagnosing) transactions of two enclosed transaction kinds are initiated: TK02 and TK03. The initiation of a transaction TK02 is optional, as expressed by the cardinality range 0..1 next to the response link from (TK01/pm) to [TK02/rq]. There may be several transactions TK03 enclosed in a TK01, as expressed by the cardinality range 1..* next to the response link from (TK01/pm) to [TK03/rq]. Fig. 7 illustrates the BPMN diagram generated from Fig. 6.

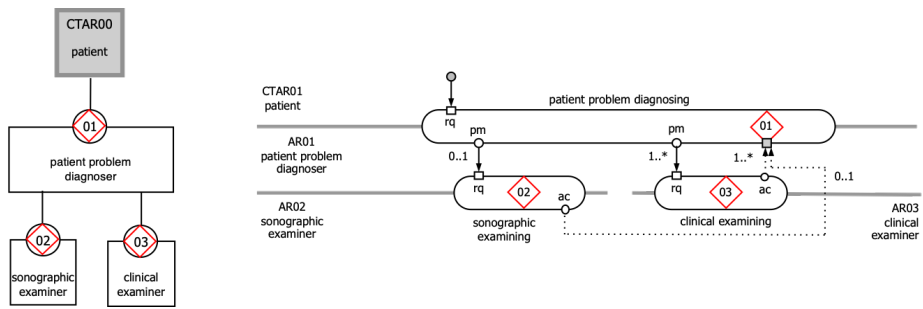

**Fig. 6. CSD and PSD of the case PoliGyn**

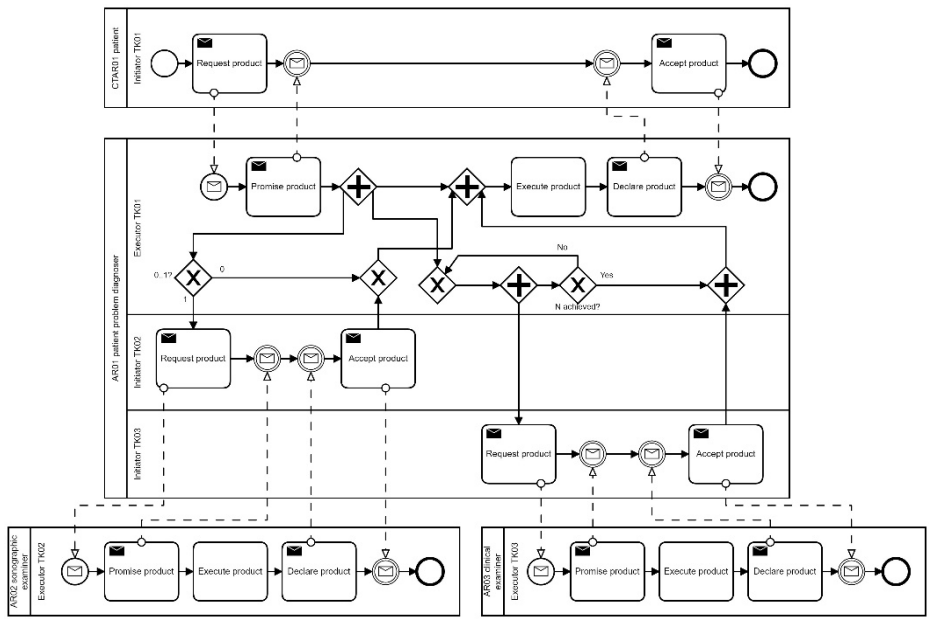

**Fig. 7. BPMN diagram of the case PoliGyn**

## 6 Reflections and conclusions

If one's understanding of business processes is not richer than their being sequences of actions and events, modelling techniques like Flowchart[2] suffice to express them in diagrams. If one wants to go one step further and include the actors who perform the actions, then EPC[3] or BPMN will do. But then the relevant question arises whether the being performed of the actions by human actors is sufficient reason to call such processes 'business processes'.

The incentive of our research is a very practical one: BPMN is the de facto standard in business process modelling. Instead of rowing against the flow, it may be wiser to join the massive community going downstream, and to try to nudge them towards embracing a richer notion of business process. Fortunately, we are not the only advocates of this approach. Comparable undertakings can be found in [4], [8], and [9].

In this short paper, we have introduced DEMO as a methodology that incorporates a notion of business process that is both very appropriate and rooted in a solid theoretical foundation [1]. We have also presented the expression of the DEMO transaction patterns, which are the building blocks of business processes, in BPMN. This appears to be quite straightforward, even for the Complete Transaction Pattern, which we only left out for saving space [6]. By including these patterns in BPMN tools, their use would be as simple as the use of the BPMN symbols. Constructing proper models of 'real' business processes, i.e. constructing trees of building blocks, is less simple but certainly doable, as we have demonstrated for the case PoliGyn from [1] in Sect. 5.

---

[2] https://en.wikipedia.org/wiki/Flowchart
[3] https://en.wikipedia.org/wiki/Event-driven_process_chain